\documentclass[12pt]{JHEP}

\usepackage{amsmath,epsfig}
\usepackage{amssymb,amsfonts}
\usepackage{latexsym}
\usepackage[vcentermath,enableskew]{youngtab}
\relax

\def\be{\begin{equation}}
\def\ee{\end{equation}}
\def\bea{\begin{eqnarray}}
\def\eea{\end{eqnarray}}
\def\hri#1#2{\href{http://arxiv.org/abs/#1}{ArXiv:#1 #2}}
\def\hre#1#2{\href{http://arxiv.org/abs/#1/#2}{[ArXiv:#1/#2]}}
\def\hspi#1#2{\href{http://www.slac.stanford.edu/spires/find/hep/www?irn=#1}{#2}}

\newcommand\fverb{\setbox\pippobox=\hbox\bgroup\verb}
\newcommand\fverbdo{\egroup\medskip\noindent%
                        \fbox{\unhbox\pippobox}\ }
\newcommand\fverbit{\egroup\item[\fbox{\unhbox\pippobox}]}

\newcommand{\la}{\lambda}

\newcommand{\bear}{\begin{eqnarray}}

\newcommand{\eear}{\end{eqnarray}}

\newbox\pippobox

\def\dc{\delta c}
\def\ls{\ell_s}

\def\6{\partial}

\def\a{\alpha}

\def\pa{\partial}

\def\C0{{\bf C_0}}
\def\Y0{{\bf Y_0}}
\def\e{\epsilon}
\def\m{\mu}
\def\n{\nu}
\def\r{\rho}
\def\s{\sigma}

\def\sp{\;\;\;,\;\;\;}

\def\a{\alpha}
\def\b{\beta}
\def\l{\lambda}

\def\ls{\ell_{\rm s}}
\def\la{\ell_{\rm AdS}}
\def\lo{\log(\Lambda r)}

\title{Dissecting the string theory dual of QCD}
\author{
\href{http://hep.physics.uoc.gr/~kiritsis/}{\large Elias Kiritsis}\\
~\\
\href{http://hep.physics.uoc.gr/}{Department of Physics, University of Crete
71003 Heraklion, Greece}\\
}

\preprint{0901.1772[hep-th]}      

\abstract{Input  from QCD and string theory is used in order to elucidate basic features of the string theory dual of QCD,
It is argued that the relevant string theory is a five-dimensional version of the type-0 superstring. The vacuum solution is asymptotically
AdS$_5$, and the geometry near the boundary is stringy. The structure of YM perturbation theory however emerges near the boundary.
In the IR,  the theory is argued to be well-approximated by a two-derivative truncation that takes into account strong coupling effects.
This explains the success of previously proposed five-dimensional Eistein-dilaton gravity with an appropriate potential
to describe salient features of the strong YM dynamics.\\
~\\
~\\
~\\
~\\
{\tt  Based on presentations made by the author in the Moriond\\
 workshop on ``Electroweak interactions and gauge theories", the workshop
 ``String Theory – From Theory to Experiment", Jerusalem,
the ``IPM String school and Workshop", Isfahan, Iran, the conference
``Strong Coupling: from Lattice to AdS/CFT" at the Galileo Galilei Institute,
Florence, the ``4th RTN EU workshop, Varna, Bulgaria
and at the conference   ``Superstrings @ Cyprus" Agia Napa, Cyprus.
It will appear in the proceedings of the Varna conference.}}

\begin{document}

\def\g{\gamma}
\def\go{\g_{00}}
\def\gi{\g_{ii}}

\maketitle 

\section{Introduction and outlook}

QCD is a very successful theory of strong interactions. It is also a theory that is hard to
 calculate with, due to the strong coupling region in the IR.
As any kind of observable  physics passes via low-energy filters it has complicated efforts in the past
 three decades to test the theory and make predictions.
Our theoretical understanding of QCD stems from several sources/techniques.

\begin{itemize}

\item  Weak-coupling perturbation theory. This is valid in the UV, because of asymptotic freedom and has been a key element
 in the understanding of the strong force.
Its impact is relying of the factorization of processes into hard and soft components.

\item  Euclidean Lattice techniques based on numerical estimates of the QCD path integral. To date this is the most direct non-perturbative technique that has
provided the first clues to confinement, and numbers for the hadron spectrum that could be compared with data, among other things.
This is a non-perturbative approach that is fully mature and its main technical limitation today is computing power.
As it is inherently Euclidean it cannot however address ab initio a class of problems that describe time-dependent dynamics.
This includes finite temperature dynamical phenomena, as well as scattering. Although some quantities can be obtained by analytic continuation
extra input is needed in order for analytic continuation to be performed reliably.

\item Special purpose phenomenological models and approximations. For specific problems, phenomenological models can give deep insights into physics that
is not directly under analytical control in QCD.
One notable example is Chiral Perturbation Theory. This is a low-energy effective field theory for the light meson sector based on ideas of chiral symmetry breaking.
Other examples include resumations of perturbative effects based on some assumptions, truncation and solutions of Schwinger-Dyson equations, applications of the
Large-$N_c$ expansion and associated matrix models, etc.

\end{itemize}

In the past decade there have been two developments that stirred the field of strong-interaction physics.
The first is data from the RHIC collider that gave the first solid indications for the physics of the quark-gluon plasma
\cite{rhic}. The second is new intuition and results on the large-$N_c$ expansion of gauge theories  that changed
our perception of the description of strongly-coupled large-$N_c$ gauge theories, \cite{malda,witten3,gkp}.
The prototype example has been the AdS/CFT correspondence as exemplified by the
(well studied by now) duality of ${\cal N}=4$ super Yang-Mills theory and IIB string theory on $AdS_5\times S^5$.
Further studies focused on providing examples that are closer to real world QCD,
\cite{D4,mnks}. It is fair to say that we now have a good
holographic understanding of phenomena
like confinement, chiral symmetry and its breaking as well as several
related issues.  The finite temperature dynamics of gauge theories,
has a natural holographic counterpart in the thermodynamics of  black-holes
on the gravity side, and the thermal properties of various holographic constructions
have been widely studied, \cite{D4,bh1,weiz,herzog,erdm}, exhibiting
the holographic version of deconfinement and chiral restoration transitions.

The simplest top-down string theory model of QCD involves $D_4$ branes with supersymmetry breaking
boundary conditions for fermions \cite{D4},
as well as a flavor sector that involves pairs of $D_8-\overline D_8$ probe branes inserted in the bulk,
 \cite{sas}. The qualitative thermal properties of this model
closely mimic what we expect in QCD, \cite{weiz}.
Although such theories reproduced the qualitative features of IR QCD dynamics,
they contain Kaluza-Klein  modes, not expected in QCD,
with KK masses of the same order as the dynamical scale
of the gauge theory. Above this scale the theories deviate from QCD.
Therefore, although the qualitative features of the relevant phenomena are correct,
 a quantitative matching to real QCD is difficult.

Despite the hostile environment  of  non-critical theory,
several attempts have been made  to
understand holographic physics in lower dimensions
 in order to avoid the KK contamination,
based on two-derivative gravitational actions, \cite{ks,bcckp}.
Indeed, large N QCD is expected to be described by a 5-dimensional theory.
The alternative problem
in non-critical theories is that curvatures are of string scale size and the truncation of the theory
to the zero mode sector is subtle and may be misleading.

 A different and more phenomenological bottom-up approach was developed and is now known as AdS/QCD.
The original idea described  in \cite{ps} was successfully applied to the meson
sector in \cite{adsqcd1}, and its thermodynamics was analyzed in \cite{herzog}.
The bulk gravitational background consists of a slice of AdS$_5$, and a constant dilaton.
There is a UV and an IR cutoff. The confining IR physics is imposed by boundary
conditions at the IR boundary.
This approach, although crude,  has
been partly successful in studying meson
physics, despite the fact that the dynamics driving chiral symmetry breaking must be
imposed by hand via IR boundary conditions.
Its shortcomings however include a glueball spectrum that does not fit very well the lattice data,
the fact that magnetic quarks are confined instead of screened, and asymptotic Regge trajectories for
glueballs and mesons that are quadratic instead of linear.

A phenomenological fix of the last problem was suggested by
introducing a soft IR wall, \cite{soft}.
Although this fixes the asymptotic spectrum of mesons and meson
dynamics
 is in principle self-consistent, it does not allow a consistent treatment
of the glue sector both at zero and finite temperature.
In particular, neither dilaton nor metric  equations of motion are solved.
Therefore the ``on-shell'' action is not really on-shell. The entropy computed from the BH horizon does
not match the entropy calculated using standard thermodynamics from the
free energy computed from the action, etc.

A well-motivated way to obtain linear Regge trajectories for mesons was advocated in \cite{paredes}.
In particular it was pointed out that the natural order parameter for chiral symmetry breaking in the context of flavor branes
is the open-string tachyon. By studying the tachyon dynamics,  it was shown that confinement in a wide class of backgrounds
is enough to guarantee chiral symmetry breaking, with linear meson trajectories.
When flavor branes are at distances larger than the string scale, the analogue of the tachyon order parameter was investigated in \cite{ak}.
This is relevant for flavor sectors that resemble more the Sakai-Sugimoto setup.

An improved holographic model that lies somewhere between bottom up and top down approaches has been proposed, \cite{ihqcd1,ihqcd2}.
It is a five-dimensional Einstein dilaton system, with an appropriately chosen dilaton potential.
The vacuum solution involves an asymptotically logarithmically AdS solution near the boundary.
The bulk  field $\l$, dual to the 't Hooft coupling,
is vanishing logarithmically near the boundary in order to match the expected QCD behavior.
This implies that the potential must have a regular Taylor expansion as $\l\to 0$, and that $\l=0$ is not an extremum of the potential.
This is unlike almost all asymptotically AdS solutions discussed so far in the literature.
In particular the canonically normalized scalar (the dilaton) is diverging at the boundary $r\to 0$ as $\phi\sim -\log(-\log r)$.
The coefficients of the UV Taylor expansion of the potential are in one-to-one correspondence with the holographic $\beta$-function.

In the IR, the potential must have an appropriate behavior so that the theory is confined, has a mass gap and a discrete spectrum.
This selects a narrow range of asymptotics that roughly obey
\be
V(\l)\sim \l^{2Q}\sp \l\to \infty~~~~~{\rm  with}~~~~~ {2\over 3}\leq Q< {4\over 3}.
\label{1}\ee
The vacuum solution always ends in a naked singularity in the bulk. Demanding that this is a ``good''
singularity in the classification of Gubser
\cite{gubserbad} implies  $Q<4/3$.
Simple interpolations between the UV and IR asymptotics reproduce very well the low-lying glueball spectrum as well as the perturbative running of the
't Hooft coupling \cite{ihqcd2}.
At finite temperature this model exhibits the behavior expected from QCD: There is a deconfining transition and the thermodynamics is very close
to what one expects from lattice QCD, \cite{ihqcd3,ihqcd4}.

In this paper we will go through several arguments originating both in string theory and QCD as we understand it, that will help us
analyse in more detail the structure of the string theory dual of QCD.
We will see that although there are ambiguities in several places, a picture emerges that seems consistent and gives some hope that we
may one day tame the associated string theory.  Even today, it may be used as qualitative litmus test of ambitious holographic models.

There are several directions that have not yet been explored.
An important one concerns the behavior of one and two-point functions in the IR.
This is an important issue as it stands at the heart of justifying the neglect of vevs of higher-dimension operators in a holographic context.
Although techniques similar to what we use can be used in this direction, we will not attempt this here.

Another issue is the theoretical definition and practical viability of a hybrid model for QCD. In such a model, physics in the UV is described via perturbative QCD
that is used to generate boundary conditions, at a rather low scale (in the few GeV region). Below this scale a holographic model should be used.
IN such a hybrid model, the UV region near the boundary (that as we argue here is stringy) can be altogether avoided.
The IR region (that as we argue here can be reasonably well-described by a two derivative action) can be handled with standard holographic techniques.
An attempt in this direction can be found in \cite{evans}.

An interesting issue is the cosmological evolution of strongly coupled matter, both made of glue and quarks.
In the former case the setup is almost identical to the one studied in the context of Randall-Sundrum cosmology as was shown in \cite{cosmo}.
Indeed the simple solution for conformal matter described in \cite{cosmo} has an alternative description in terms of lowering the UV cutoff brane inside AdS.
The Randall-Sundrum tuning corresponds to the choice of coupling gravity to renormalized rather than bare sYM operators (the vacuum energy in particular).
A similar study for a non-conformal theory like QCD involves a few extra ingredients, the most important of which is establishing  the geodesic
motion of the UV boundary in the bulk, and in particular the dilaton couplings to the boundary. This is interesting as it may give new tools to study the
impact of the deconfinement phase transition in a cosmological setup.

\section{General remarks on the string theory dual\label{general}}

The first  question we may pose is: in how many dimensions is the string theory dual of QCD  living?
A way to answer this question proceeds via the intuition developed in the past 20 years from matrix and other large $N$ theory duals
to string theory. Indeed, the intuition is as follows. The large N-gauge theory contains several adjoint fields living on a d-dimensional space $M_d$.
Typically the eigenvalues of the adjoint matrices becomes new continuous dimensions. For example in the case of the ``old matrix models"
the single eigenvalue distribution increases the spacetime dimension by one.

Not all adjoint fields provide independent eigenvalue distributions and therefore holographic dimensions.
Fields related by global symmetries must be reduced appropriately. Also if the symmetry is not there, but there is a symmetric theory related by RG flow to the previous one,
than there is a reduction in the number of holographic dimensions.
For example in ${\cal N}=4$ superYM, there are 4 adjoint matrices from the vectors, 8 adjoint matrices from the fermions
 and 6 adjoint matrices from the scalars. Four-dimensional Lorentz invariance of the vacuum indicates there is a single
  independent eigenvalue distribution from the vectors. Similarly the SO(6) symmetry implies that that there are 5 independent eigenvalue distributions
  from the scalars (nicely exposed in the work of Berenstein and collaborators, \cite{bere})

In the case of QCD the situation is simpler. We have four (a vector) adjoint fields. Since the theory and its vacuum on flat space are expected
to be Lorentz invariant, only one eigenvalue distribution is independent. Therefore we expect one extra holographic dimension and therefore the
string theory dual is expected to live in 5 dimensions.

QCD unlike successful gauge theories in the holographic domain, is asymptotically free. This is a property that has marred attempts
 in the past 10 years to work out a trustworthy dual string theory. Intuition coming from ${\cal N}=4$ superYM indicates that
 the 't Hooft coupling is directly related to the  spacetime curvature as $\lambda\sim {\ell^4_{\rm AdS}\over \ell^4_{\rm s} }$.
If this is taken at face value, it would imply that the putative QCD dual will have singular curvatures near the UV, where the QCD
coupling vanishes.

There are several caveats to this line of reasoning, that we will mention here.
\begin{itemize}

\item The relation between the 't Hooft coupling and spacetime curvature as implied by ${\cal N}=4$ superYM may not be universally applicable
and in particular its extrapolation to weak coupling may not be trustworthy.
We have many examples of such behavior, where non-linear behavior of effective actions
(notably the DBI and CFT actions) smooths out the leading singular behavior. Controlable examples in closed string theory itself, that include WZW and coset models
do indeed behave differently at strong curvature. A suggestive example is SU(2)$_k$ WZW model where the limit of strong curvature has either curvature of
order the string scale (k=1) or contains no space at all (k=0).

It is plausible that  a respectable alternative is that in the limit of the vanishing coupling,  curvatures remain at the string scale.
Therefore although the regime is stringy, it is not singular.

\item  The 't Hooft coupling in ${\cal N}=4$ is constant. As such, one can rescale the metric and compute the curvature in different ``frames".
Although different rescaled curvatures behave differently, going from one to the other is a simple process.
On the other hand the choice of frame is a relevant
question in the string theory dual of QCD, where $\l$ is expected to be a function of the holographic coordinate, reproducing the RG running of the YM coupling.
Different frames with include derivatives of the $\l$ in the curvature and can radically alter its behavior.

\item Another issue is the approximate conformal invariance that characterizes QCD in the extreme UV. We do expect that this will be geometrically
 encoded in the string theory dual. The generic lesson of holography in controlled examples is that conformal invariance and its approximations are
 encoded in the structure of the string geometry via AdS spaces and their deformations.
 AdS is a very symmetric spacetime, and carries a single boundary where the UV definition of the holographic theory resides via sources.
 It would be a rather unfortunate situation that the space is singular at the place where we are supposed to define the UV of the theory.

\end{itemize}

Because of all of the above, a reasonable expectation of the structure of the string
background describing the QCD dual is a metric that is asymptotically
$AdS_5$ near its boundary region (UV). We expect that this will be modified as we flow
 towards the IR signaling the breaking of conformal invariance.
Moreover as we will argue more in detail further on, the AdS$_5$ curvature near the
boundary should be comparable to the fundamental string length $\ell_{\rm s}$.

\subsection{The low-energy string spectrum: a gauge theory view}

An important question that is crucial in setting up the vacuum problem in the string theory dual is to estimate
which of the bulk fields are important in determining the vacuum structure.
The lowest dimension fields are certainly the most important in the UV.
QCD has no (strongly) relevant operators. The first non-trivial operators  start at dimension 4
  and are given by the quadratic trace of the field strengths
\be
Tr[F_{\m\n}F_{\r\s}]
 \label{1}\ee
To distinguish different operators we may use the U(d) decomposition

\be
\left(\Yboxdim8pt\yng(1,1)\otimes \Yboxdim8pt\yng(1,1)\right)_{\rm symmetric}
 =\Yboxdim8pt\yng(2,2)\oplus \Yboxdim8pt\yng(1,1,1,1)
 \label{2}\ee
Finally we must remove traces to construct the irreducible representations of O(d):
\be
\Yboxdim8pt\yng(2,2)=\underline{\Yboxdim8pt\yng(2,2)}\oplus \underline{\Yboxdim8pt\yng(2)} \oplus \bullet
\sp  \Yboxdim8pt\yng(1,1,1,1)= \bullet
 \label{3}\ee
where the line under a Young tableau implies that all possible traces have been removed and it therefore represents an irreducible representation
of O(d). A $\bullet$ stands for the singlet of O(d).

The two singlets  are  the scalar (YM Lagrangian, dual to the string theory dilaton)  and pseudoscalar (instanton) densities:
\be
\phi\leftrightarrow Tr[F^2]\sp a\leftrightarrow Tr[F\wedge F]
 \label{4}\ee
Each carries a single d.o.f.
The t' Hooft coupling $\l$ is related to the dilaton as
\be
\l\sim N_ce^{\phi}
\ee
at least in the UV.
The instanton density should be dual to a pseudoscalar bulk axion, $a$.
 The next operator is a traceless conserved symmetric tensor
\be
\underline{\Yboxdim8pt\yng(2)}~~\to~~ T_{\m\n}=Tr\left[F^2_{\m\n}-{1\over 4}g_{\m\n}F^2\right]\sp {(T)_{\m}}^{\m}=0\sp \pa^{\m}T_{\m\n}+0
 \label{5}\ee
 It is dual to the 5-dimensional graviton, $g_{\m\n}$.
Finally
\be
\underline{\Yboxdim8pt\yng(2,2)}~~\to~~T^4_{\m\n;\r\s}=Tr[F_{\m\n}F_{\r\s}-{1\over 2}
(g_{\m\r}F^2_{\n\s}-g_{\n\r}F^2_{\m\s}-g_{\m\s}F^2_{\n\r}+g_{\n\s}
F^2_{\m\r})+{1\over 6}(g_{\m\r}g_{\n\s}-g_{\n\r}g_{\m\s})F^2]
 \label{6}\ee
It has 10 independent d.o.f and  should be dual to a similar massive tensor in the bulk string theory.

Unlike ${\cal N}=4$ superYM, near the UV, the dimensions of these operators are reliable,
as they are given by their free-field theory values plus small corrections.
Of all these operators with $\Delta=4$, only the last one correspond to a field that is massive in the bulk string theory. Therefore it is expected to be
less important at least in the UV.

Operators with $\Delta=5$ are given by $Tr[\nabla_{\m}F_{\n\r}F_{\s\tau}]$ (with $\nabla$ being the gauge covariant derivative)
while operators with $\Delta=6$ are given by
$Tr[\nabla_{\m}F_{\n\r}\nabla_{\m'}F_{\s\tau}]$ and $Tr[F_{\m\n}F_{\m'\n'}F_{\r\s}]$.
Of all the higher dimension operators one is expected to correspond to the NS antisymmetric tensor,
 namely
 \be
 B_{\m\n}\sim   Tr[F_{[\m a}F^{ab}F_{b\n]}+{1\over 4}F_{ab}F^{ab}F_{\m\n}]
 \ee
  \cite{f3}
  and should be massive with a UV mass $M={4\over \ell_{\rm AdS}}$ in order to have the correct scaling dimension in the UV.
   This is happening because of a combination of two effects : the fact that
  $B$ appears in the five-form field strength together with the RR two-form $C_2$ and the fact that the RR-five form has a vev. This provides a mixing
  of $B_2$ and $C_2$ that gives both a mass. This is similar to what happens in ${\cal N}=4$ superYM.

The conclusion of this section  is that the massless bulk fields should be dual
 to the operators, $Tr[F^2]$ (dilaton), $T_{\m\n}$ (metric), $Tr[F\wedge F]$ (axion).

\subsection{Bosonic string or superstring?\label{axion}}

We will now present an argument that suggests that the string theory dual to QCD should
 have a RR sector, which furthermore implies that it is a superstring theory.
This does not necessarily imply (broken) spacetime supersymmetry but rather that the
world-sheet gauge symmetry of the theory is some form of supersymmetry, which implies
in particular the existence of a (bi-fermionic), RR sector.
Note however that pure YM does not have gauge invariant fermionic operators.
Therefore, the string theory, although a superstring theory, it should not contain
spacetime fermions (NS-R, and R-NS sectors).
Type 0 theories, have precisely this property, and have been candidate dual grounds
 for YM for some time \cite{type0}.

It should be noted here that even after we add quarks, all gauge invariant operators
 in QCD (with the exception of baryon operators) are bosonic.
Baryon operators on the other hand should correspond to appropriate ``solitonic" D-branes
in the string theory as we now understand in many similar
examples \cite{witten2}. Therefore the standard spectrum of the string theory dual to QCD should also
 contain no spacetime fermions.

 A first candidate for a RR field should be the RR four-form, $C_4$, that in standard examples
 is known to provide the flux responsible for introducing the
 (large number of ) $D_3$ branes into the background geometry. Five dimensions is a very
  special dimension however for $C_4$ as it does not contain
 any propagating degrees of freedom, and this makes  therefore its presence a bit murkier.
There is a however another bulk field that (a) is propagating and (b) sources one of the couplings of the YM theory.
This is the  axion, dual to the YM instanton density. This field must be a RR field (as
indeed happens in the ${\cal N}=4$ superYM example),
in order to match  known properties of the CP-odd sector of large-$N_c$ YM \cite{witten}.

The action of large-$N_c$ QCD including a $\theta$ angle can be written in the form
\be
S_{YM}=\int d^4x~Tr\left[{1\over 4g^2}F^2+{\theta\over 8\pi^2}F\wedge F\right]=N_c
\int d^4x~Tr\left[{1\over 4g^2N_c}F^2+{\theta\over 8\pi^2N_c}F\wedge F\right]
 \label{7}\ee
The proper scaling in the 't Hooft large-$N_c$ limit is to keep
$\l=g^2N_c$ and $\zeta={\theta\over N_c}$ finite and fixed.
Consider now the $\theta$-dependent vacuum energy, to leading order
in the $1/N_c$ expansion
\be
E_{YM}=N_c^2 F[\lambda,\zeta]
 \label{6}\ee
This must be invariant under the $\theta$-angle periodicity shift
$\theta\to\theta+2\pi$. This however is impossible if F is a smooth function.
This can be achieved only  if $F$ is multibranched function (obtained by minimizing
 a collection of many nearly degenerate minima).

In view of this we can write the vacuum energy as \cite{witten},
\be
E_{YM}=N_c^2 Min_{k\in Z}f\left[\lambda,{\theta+2\pi k\over N_c}\right]
 \label{7}\ee
which shows that it is periodic but it is not a continuous function of $\theta$.
The CP transformation $\theta\to -\theta$ implies that $f(\l,\theta)=f(\l,-\theta)$.
CP is unbroken only if $\theta=0,\pi$.

The integer $k$ labels different vacua that are related by integer shifts of the $\theta$- angle. The absolute minimum
is expected at $\theta=0$ at the $k=0$ vacuum. Taking all this into account we can write at  large $N_c$
 \be
E_{YM}=N_c^2 f\left[\lambda,0\right]  +{1\over 2}{\pa f\left[\lambda,0\right]\over \pa \zeta}
Min_{k\in Z}~(\theta+2\pi k)^2+{\cal O}(N_c^{-2})
 \label{8}\ee
The quantity $\chi={\pa f\over \pa \zeta}\left[\lambda,0\right]$
is known as the topological susceptibility and it is known to be non-zero
\cite{treview}. We therefore observe that the leading $\theta$ dependence of the vacuum energy is coming in at order ${\cal O}(1)$
while higher terms are further suppressed with $N_c$.

We will now show that this property on the string theory side is due to the special properties of RR fields.
The axion is dual to the instanton density. This implies that its source  gives the UV value of the $\theta$ angle
\be
a(r)=\theta_{UV}+{\cal O}(r^4)
 \label{9}\ee
Assuming it is a RR field we may write its tree-level effective action as
\be
S=M^3\int d^5x\sqrt{g}e^{-2\phi}\left[R+4(\pa\phi)^2+{e^{2\phi}\over 2}(\pa a)^2+C_4~e^{4\phi}(\pa a)^4+\cdots\right]
 \label{10}\ee
where it should be noted the peculiar dependence on the dilaton of the axion terms
in agreement with standard string-theory dilaton counting.
We now translate to variables that have a smooth large-$N_c$ limit $\l=N_c e^{\phi}$ to rewrite the action with explicit $N_c$ dependence
\be
S=N_c^2~M^3\int d^5x\sqrt{g}{1\over \l^2}\left[R+4\left({\pa\l\over \l}\right)^2+{\l^2\over 2N_c^2}(\pa a)^2+C_4~{\l^4\over N_c^4}(\pa a)^4+\cdots\right]
 \label{11}\ee
As in the on-shell action $(\pa a)^2\sim \theta^2$, $(\pa a)^4\sim \theta^4$ etc, we observe that we obtain the same large-$N_c$ scaling of
the different $\theta$-dependent terms as in the field theory side , (\ref{8}). More details on the vacuum solutions and action for the QCD axion can be found in
\cite{ihqcd2}.

The upshot of the previous analysis is that the axion in the string theory dual of QCD is a RR field, indicating
that the string theory is a superstring theory in the type-0 class.

\subsection{The minimal low-energy spectrum: a string-theory view}

From our discussion so far we have argued that the string theory dual must have a NS-NS sector
with the usual fields, $g_{\m\n}, B_{\m\n},\phi$, as well as
a RR sector that contains at least the axion $a\equiv C_0$ and
the four-form gauge potential $C_4$ necessary for generating the color flux, which will be proportional to the
number of colors $N_c$.

Two main issues require a discussion. The first is the RR sector. The minimal possibility is a spinor$\times$ spinor in five dimensions.
Five-dimensional spinors are not chiral, therefore no gaps are expected in the expansion of the RR bispinor.
A direct group-theoretic expansion gives
\be
{\rm spinor}_5 \times {\rm  spinor}_5= F_0+F_1+F_2+F_3+F_4+F_5
 \label{12}\ee
where $F_p=dC_{p-1}$ is a p-form field strength.
 Moreover, the truly independent fields, are $F_{0,1,2}$ as the rest are related by Poincar\'e duality: $F_5=~^*F_0$, $F_4=~^*F_1$, $F_3
 =~^*F_2$ as implied by the properties of spinors.

 \begin{itemize}

 \item $F_5=~^*F_0$. $F_5\sim N_c$ and the associated $C_4$ is not a propagating field in five dimensions. Therefore the 5-flux is important as a background, and
 as argued in \cite{ihqcd1,ihqcd2} is responsible for non-trivial dilaton dependence in the tree-level string effective action.
 It provides an IR effective potential for the dilaton (=QCD coupling), and other important dynamical effects in the flavor sector, in particular the
 correct flavor anomaly generating CS terms \cite{paredes, ihqcd2}. Having no propagating degree of freedom, it is not dual to any operator of large-$N_c$
 YM.

 \item $F_4=~^*F_1$. $F_1$ is the field strength of the axion $a$ dual to the QCD instanton density. Its dual form is a three-form $C_3$.

 \item  $F_3=~^*F_2$. $F_3$ is the field strength of a RR two-form, $C_2$. Its dual is a one-form $C_1$. Although these forms belong to the lower level
  of the RR spectrum, they should correspond to massive states of the string theory, and therefore to higher (than four) dimension operators in QCD.
 $C_2$ becomes massive due to its (CS related) mixing with $B_2$ from the NS-NS sector. Ignoring the axion we may write the leading order (string-frame)
 action for the three- and five-forms
 as
 \be
 S=-M^3\int d^5x\sqrt{g}\left[{e^{-2\phi}\over 2\cdot 3!}H_3^2  +{1\over 2\cdot 3!}F_3^2+{1\over 2\cdot 5!}F_5^2\right]
  \label{13}\ee
 \be
 F_3=dC_2\sp H_3=dB_2\sp F_5=dC_4-C_2\wedge H_3
  \label{14}\ee
 The equations of motion that stem from this action are\footnote{These equations are consistent only if the dilaton is constant, which is the case in
 the ${\cal N}=4$  sYM case. Here non-linearities are important as we will see later one, but the basic mixing mechanism is similar.}
 \be
 \nabla^{\m}(e^{-2\phi}H_{3,\m\n\r})+{1\over 4}F_{5,\n\r\a\b\g}{F_{3}}^{\a\b\g}=0
 \sp \nabla^{\m}F_{3,\m\n\r}+{1\over 4}F_{5,\n\r\a\b\g}{H_{3}}^{\a\b\g}=0
  \label{15}\ee
 \be
 \nabla^{\m}F_{5,\m\n\r\s\tau}=0~~~\to~~~F_{5,\m\n\r\s\tau}={\e_{\m\n\r\s\tau}\over \sqrt{g}}{w~N_c\over \ls}
  \label{16}\ee
 where $w$ is a dimensionless constant.
 Substituting the five-form in the three-form equations we obtain,
\be
 \nabla^{\m}(e^{-2\phi}H_{3,\m\n\r})+{wN_c\over 4\ls}{\e_{\n\r\a\b\g}\over \sqrt{g}}{F_{3}}^{\a\b\g}=0
 \sp \nabla^{\m}F_{3,\m\n\r}+{wN_c\over 4\ls}{\e_{\n\r\a\b\g}\over \sqrt{g}}{H_{3}}^{\a\b\g}=0
  \label{17}\ee
 We may decouple the two equations by direct manipulation to obtain
 \be
 \nabla^{\m}\left[\nabla^{\n}(e^{-2\phi}H_{3,\m\r\s}+{\rm cyclic}\right]+{3w^2N_c^2\over 16\cdot 5!\ls^2}H_{3,\nu\r\s}=0
  \label{18}\ee
 and a similar one for $F_3$.
 The $N_c$ dependence is also appropriate as the equation (\ref{18}) is $N_c$-independent if we introduce the appropriate $N_c$-independent variable
 $\l=N_c e^{\phi}$ proportional to the YM 't Hooft coupling in the UV.

 The upshot of this analysis is that both $B_2$ and $C_2$ combine to a massive two-tensor, that should be dual to the $C-odd$ non-conserved operator
  $Tr[F_{[\m a}F^{ab}F_{b\n]}+{1\over 4}F_{ab}F^{ab}F_{\m\n}]$ with UV dimension 6.

 \end{itemize}

Therefore, this minimal spectrum includes all fields we expect to be massless in five dimensions ($g_{\m\n}, \phi,a$) as well as the
flux-generating four-form.
In simple type-0 vacua in ten dimensions, there is a doubling of RR fields. This is due to the fact that there is effectively no chirality projection.
However, in 5 dimensions, there is no chirality and it is expected that the RR sector would have the form advocated above.

 There is another point that needs discussion: in type-0 vacua in ten and six
 dimensions there is a closed-string ``tachyon" scalar. It is indeed a tachyon
 in flat ten-dimensional space \cite{type1}, but it may be massless or even
 massive in curved non-critical backgrounds \cite{niarchos}.
 There is certainly no place for a tachyonic or massless scalar near the boundary
  of AdS in the dual string theory of YM as that would imply the presence of
 another relevant or marginal operator in YM. However all such operators have already accounted for.
 Therefore, if the string theory has such a zero-th level scalar, it should be
 massive. Moreover, it is not at all obvious what operator should be the dual
 of such a scalar\footnote{Some intriguing observations on the couplings of such a scalar to probe D-branes were made in \cite{niarchos}.}.

Of the minimum set of fields that we mentioned above, only $g_{\m\n},\phi,a, F_5$ can have vevs (non-trvial profiles) in the vacuum in order to preserve O(1,3)
Lorentz invariance. In particular this precludes vectors and two-index antisymmetric tensors from obtaining a vev.

\subsection{The relevant charged defects}

Several strings and branes are expected to exist as solitonic objects in this string theory, in analogy with critical string theory.
We will enumerate them trying to elucidate the nature of each defect.

\begin{itemize}

\item {\tt The fundamental string}. It couples electrically to $B_{\m\n}$. This is expected to represent the YM flux tube.
  Its tension, $T_{F}\sim {1\over \ls^2}$ is $N_c$-independent. It should not be confused
with the QCD string tension $\s$ that is multiplying the linear term in the
 inter-quark potential. This is proportional to $T_F$ but the string-frame scale factor also enters
\cite{ihqcd2}.

\item {\tt $NS_0$ brane}. In five dimensions it couples magnetically to $B_{\m\n}$.
It is the analogue of the $NS_5$ brane of critical string theory.
It is a ``point-like" soliton with tension that scales as $T_{NS_0}\sim {\cal O}(N_c^2)$.
It should be thought of as a magnetic baryon vertex that binds together $N_c$
magnetic quarks (each having a mass that scales as ${\cal O}(N_c)$) .

\item {\tt $D_{-1}$ branes}. They are the YM instantons.
They couple electrically to the axion. Their ``tension" is ${\cal O}(N_c)$ in agreement with what we expect from the instanton action.

\item {\tt $D_{0}$ branes}.  They couple magnetically to $C_2$. They are the baryon vertices in five dimensions.
Their WZ couplings are responsible for binding $N_c$  fundamental strings.
Their tension is ${\cal O}(N_c)$. On flavor branes they couple to
 baryon number that is equivalent to instanton number of the flavor $U(N_f)$ gauge fields.

\item {\tt $D_{1}$ branes}. They are the magnetic strings, namely flux tubes
between magnetic quarks. They couple electrically to $C_2$, and have a tension of order ${\cal O}(N_c)$.

\item {\tt $D_{2}$ branes}. They couple magnetically to the axion. They are domain walls in 4 dimensions
 that separate different oblique confinement vacua. As one moves across a $D_2$ brane, $\theta$ jumps by $2\pi$.

\item {\tt $D_{3}$ branes}. They couple electrically to the four-form. They generate the gauge group of the gauge theory.

\item {\tt $D_{4}$ branes}. They are space filling branes that generate flavor in the YM theory.
 Since the string theory is oriented, tadpole cancelation implies that
an equal number $N_f$ of $D_4$ and $\overline{D_4}$ branes must be introduced in 5 dimensions. The strings stretched between
$D_3$ and $D_4$ branes generate the left-handed quarks while the ones  stretched between
$D_3$ and $\overline{D_4}$ branes generate the right-handed quarks.
\end{itemize}

\subsection{Why a spectrum truncation might work for the vacuum structure?}

Large-$N_c$ YM has an infinite number of single-trace operators. The UV definition
 of the theory, involves only $T_{\m\n}$, and $Tr[F^2]$.
If $\theta_{UV}$ is non-zero, then $Tr[F\wedge F]$ is also involved. In the
holographic dual this translates into the statement that in the vacuum solution, only
$\phi, g_{\m\n}$ and potentially $a$ have a source term in the UV boundary. This guarantees
that their profile in the string-theory vacuum solution is necessarily non-zero.
Although all the other (infinite tower) of bulk fields have no UV sources, this does not necessarily imply
that their profiles in the vacuum solution vanish. They can have non-trivial vevs,
 that would trigger a non-trivial solution profile in the holographic direction.

 In a theory with exact conformal invariance, non-zero one-point functions, can be
 redefined to zero by a subtraction. Because the theory is conformal, once they are set
 to zero at a given scale, they remain zero at all scales.
 This is not the case in theories where conformal symmetry is broken , as in QCD.
 Although one can subtract a one-point correlator a a given scale, this does not guarantee than the
 one-point function remains zero at all scales.
 Of course, bulk fields that break Lorentz invariance cannot acquire vevs, and therefore have trivial profiles in the vacuum.
 But this is not the case for example for bulk scalars.
 One therefore may ask, can we neglect the non-trivial profiles of all other single trace operators in the YM/QCD string vacuum solution?

 The answer in the UV is rather simple: the higher the scaling dimension and spin of an operator, the larger its bulk mass, and the smaller its influence
 in the equations of motion of basic fields, $g_{\mu\n},\phi,a$.
 This is a well-known effect both in asymptotically AdS space-times and asymptotically free QFTs.
 This is the reason we can truncate the infinite coupled system of
 RG equations near a free field theory, and study a small number of flows corresponding to the most relevant operators.
 Therefore in the UV of YM, (free-field) scaling dimensions and spin determine the relative importance of operators.

 The situation in the IR is more complicated and strictly speaking beyond
 control in QCD. Generally speaking, higher-dimension operators can and do
 sometimes become important
 in theories that are strongly coupled in the IR. A typical example involves
  (massive) KK fields in the bulk that their vevs
  can be important in resolving IR singularities in the bulk.
  Is this expected to happen in QCD? The direct,  short and honest answer is :
   we do not know. There are however a few tantalizing arguments and
  indications in the past two-three decades that point towards the following answer:
   that for many, IR relevant and simple  observables, vevs of higher
  dimension (in the UV) operators are not that important for the IR physics.
  A large class of such arguments are summarized by the surprising successes
  of the SVZ sums rules \cite{shifman} that seem to hint at the previous statement.
   A holographic argument in the same direction, elaborated in the context
   of the string theory dual will also be presented later on in this article.
   In view of this we will entertain the possibility that for several observables
   it is enough to consider the vacuum structure as described by ``light string
    spectrum", $g_{\m\n}, \phi,a$ together with the the non-propagating four-form
   potential.

\subsection{The vacuum solution ansatz}

We will consider YM and QCD on Minkowski space, or Euclidean four-dimensional (flat) space.
 More exotic geometries maybe considered
but we will not explore them here to keep the discussion simple. The theory is Lorentz invariant and
 we do not expect that Lorentz invariance will be broken
in the quantum theory.
Therefore the form that the metric and other fields will take in the vacuum is very constrained
\be
ds^2\equiv g_{\m\n}dx^{\m}dx^{\n}=b(r)^2[dr^2+dx^{M}dx_{M}]\sp \phi\to \phi(r)\sp a\to a(r)
\label{g}\ee
where $g_{\m\n}$ is the five-dimensional metric, $x^{M}$ are the four-dimensional Minkowski coordinates
and $r$ is the radial coordinate. One can still perform radial reparametrizations, which have been used to bring the metric to the
form above. There is also the five-form field strength $F_5$.  4d Lorentz invariance   implies that
\be
F_{\m_1\m_2\cdots \m_5}=f(r)~\epsilon_{\m_1\m_2\cdots \m_5}
\ee

The fact that we are in five dimensions and the fact that Lorentz-invariance constraints the metric as in (\ref{g}) already implies that
capped geometries, like the one in \cite{d4} that have been popular in order to
describe backgrounds that are confining in the IR are not possible here. The reason is that there is no extra holographic
 coordinate beyond the radial one to generate the cigar geometry familiar from Euclidean black holes.
 Lorentz invariance on the other hand prohibits the used of a Minkowski coordinate to that effect.

From now and for the rest of this paper we will neglect the axion.
As discussed in section \ref{axion}, its kinetic term is large-$N_c$ suppressed
and it does not therefore contribute to the vacuum structure at
leading order in 1/$N_c$, except for CP-odd observables. The YM axion has been discussed
in \cite{ihqcd1,ihqcd2}.

\section{The string effective action\label{eft}}

In this section in preparation for our exploration of the ``vacuum" of the string theory dual to YM
we will investigate the general action at string tree level.
As we will see, at least in the UV, the geometry is expected to be stringy, therefore we do not expect a
few terms in the derivative expansion to be a good guide.
In the absence of an exact string description we will use the language of arbitrary functionals of local curvature invariants as a guide.
Although this intuition may sometimes  fail we will trust it to obtain qualitative conclusions.

The tree-level string effective action that we will start (in the string frame) is
\be
S_{\rm tree}=M^3\int d^5x~\sqrt{g}~e^{-2\phi}\left[4(\partial \phi)^2+ F(R,\xi)\right]\sp \xi\equiv -e^{2\phi}{F^2_5\over 5!}
\label{b1}
\ee
In this action we have included the fields that will be non-trivial in the
 vacuum solution, namely $g_{\m\n}$, $\phi$, $F_5$.
We have suppressed most possible distinct tensor structures under which the
 curvature and the five-form field strength enter in the above action.
In particular, the effective action is expected to be a non-linear function
of the Riemann tensor and its covariant derivatives.
Although such terms are important ingredients in several string theory
observables\footnote{In particular the holographic conformal anomaly \cite{skenderis}
and the shear viscosity \cite{son} can depart from their universal values
only if higher derivative structures involving the Riemann tensor are present.},
for the arguments that will be made,  a simplified action involving
only the scalar curvature $R$ and the five-form square $F_5^2$ will suffice.
Note that the five-form is always accompanied by a power of $e^\phi$ as is
the case for RR forms.

The function $F$ in (\ref{b1}) will be taken arbitrary at this stage.
Notice also that we did not add non-linear terms in the kinetic term of the
 dilaton, in the string frame. Following many works and explicit analysis
of the string $\sigma$-model up to four-loops it was conjectured \cite{tseytlin}
that there is only a linear term in $(\pa \phi)^2$. We will assume this to be true here.

For weak curvatures we can expand
\be
F={2\over 3}{\dc\over \ls^2}+R+{1\over 2}\xi+{\cal O}(R^2,R\xi,\xi^2)\sp \dc=10-5=5
\label{b2}\ee
to obtain the standard two derivative tree effective action including the dilaton potential (the first term)
 present due to the fact that we are in non-critical string theory.

As the four-form potential is non-propagating in five dimensions, it is
 appropriate to ``integrate it out".
To do this we derive its equations of motion, solve them, we then substitute
back into the equations of motion of the others fields
(metric and dilaton) and we find the new action from which these equations
stem by variation.

Varying (\ref{b1}) with respect to the four-form we obtain the equation
\be
\nabla^{\m}\left(F_{\xi}~F_{\m\n\r\s\tau}\right)=0
\label{b3}\ee
where $F_{\xi}\equiv {\pa F(R,\xi)\over \pa \xi}$.
The solution of (\ref{b3}) is
\be
F_{\xi}~F_{\m\n\r\s\tau}={N_c\over \la^2}{\epsilon_{\m\n\r\s\tau}\over \sqrt{g}}
\label{b4}\ee
where called the constant $N_c$ and inserted the AdS  scale $\la$ that will
 be introduced later in (\ref{b17}) to make $N_c$ dimensionless\footnote{Although
  the AdS length will emerge later,
it is inserted  here for economy, in order to make later formulae simpler.}.
 All we know of course is that this constant is linear in the number of colors, but unlike the
critical  ${\cal N}=4$ case we do not know the precise coefficient. To obtain
this we need to know the $D_3$ solutions in the non-critical theory in question
and their tensions. We will still call the flux constant $N_c$ though from now on,
 keeping in mind that this is only {\em proportional}
to the number of colors. What is important is that we are working in the limit
where this flux is sent to infinity keeping
\be
\lambda \equiv N_c e^\phi
\ee
fixed. Indeed this combination is proportional to the 't Hooft coupling of YM, at least in the UV.
Squaring (\ref{b4}) we obtain
\be
\xi~F_{\xi}^2={\lambda^2\over \la^2}
\label{b5}\ee
This is an algebraic equation that involves, $\xi,R, \lambda$, and we must solve it implicitly
to obtain $\xi$ as a function of $R,\lambda$.
This can be used to obtain the following differential formulae,
\be
(F_{\xi}+2\xi F_{\xi\xi})d\xi+2\xi F_{\xi R}dR=2\xi F_{\xi}d\phi\sp
{d\xi\over d\phi}={2\xi F_{\xi}\over F_{\xi}+2\xi F_{\xi\xi}}
\label{b29}\ee
useful
 for the variations of the action with respect to the other fields.
The equations for the dilaton and graviton read
\be
\square \phi-(\pa\phi)^2+{1\over 4}\left(F-\xi F_{\xi}\right)=0
\label{b6}\ee
\be
F_R ~R_{\m\n}+e^{2\phi}(g_{\m\n}\square-\nabla_{\m}\nabla_{\nu})(e^{-2\phi}F_{R})+4\pa_{\m}\phi\pa_{\n}\phi-{1\over 2}(4(\partial\phi)^2+F)g_{\m\n}-
F_{\xi}{e^{2\phi}F_{\m\n}^2\over 4!}=0
\label{b7}\ee
where $F_R\equiv {\pa F(R,\xi)\over \pa R}$.
Substituting from (\ref{b5}) into (\ref{b7}) we obtain the equation
\be
F_R~R_{\m\n}+e^{2\phi}(g_{\m\n}\square-\nabla_{\m}\nabla_{\nu})(e^{-2\phi}F_{R})+4\pa_{\m}\phi\pa_{\n}\phi-
{1\over 2}(4(\partial\phi)^2+F)g_{\m\n}+{\xi~ F_{\xi}}g_{\m\n}=0
\label{b8}\ee

Equations (\ref{b6}) and (\ref{b8}) can be obtained from the following equivalent action
\be
S_{\rm tree}=\int d^5x\sqrt{g}e^{-2\phi}\left[4(\partial \phi)^2+ F(R,\xi)-2\xi F_{\xi}(R,\xi)\right]\sp
\label{b9}
\ee
where in this action $\xi\equiv \xi(R,\l)$ is an algebraic solution of (\ref{b5}).
Note that this action is ${\cal O}(N_c^2)$ if we use variables that are finite in the large-$N_c$ limit as follows
\be
S_{\rm tree}=M^3N_c^2\int d^5x\sqrt{g}~{1\over \l^2}\left[4{\partial \l^2\over \l^2}+ F(R,\xi)-2\xi F_{\xi}(R,\xi)\right]\sp
\label{b9a}
\ee

The conclusion  of this analysis is that in the end of the day we must solve the following two equations subject to the algebraic
condition (\ref{b5})
\be
F_R~R_{\m\n}+e^{2\phi}(g_{\m\n}\square-\nabla_{\m}\nabla_{\nu})(e^{-2\phi}F_{R})+4\pa_{\m}\phi\pa_{\n}\phi-
{1\over 2}(4(\partial\phi)^2+F)g_{\m\n}+\xi~ F_{\xi}g_{\m\n}=0
\label{b10}\ee
\be
4\square \phi-4(\pa\phi)^2+F-\xi F_{\xi}=0
\label{b11}\ee

\section{The UV regime\label{uv}}

It was already argued in section \ref{general} that the most reasonable description of the asymptotic UV geometry
is as an AdS$_5$ near-boundary region. This description geometrizes the asymptotic conformal invariance,
and brings us in line with what was understood in the best studied case of
${\cal N}=4$ superYM.

The conformal invariance in QCD however comes together with a coupling constant that vanishes in the UV.
Therefore we expect that as we approach the boundary $\l\to 0$.
In the metric ansatz (\ref{g}) our expectation translates into the fact that near the boundary
\be
b(r)\to {\la\over r}
\ee
the AdS$_5$ warp-factor in Poincar\'e coordinates.
Moreover in this regime the intuition is similar to the ${\cal N}=4$ case: r is serving as the inverse of the energy scale.

QCD tells us that near the UV, the 't Hooft coupling depends on the energy as
\be
{1\over \l}={b_0\log{E\over \Lambda}}+{\cal O}(\log\log{E\over \Lambda})
\ee
Using $r$ as the inverse of the energy we deduce that the solution for the 't Hooft coupling  $\l$ near the AdS boundary must look like
 \be
{1\over \l}=-{b_0\log{r\Lambda}}+{\cal O}(\log\log{r\Lambda})
\ee
and therefore $\l\to 0$ as ${1\over \log{r\Lambda}}$ as we approach the boundary $r\to 0$.

We conclude that asymptotically close to the boundary $R\to R_*=-{20\over \la^2}$ and $\l\to 0$.

Consider now equation (\ref{b5}). Since the right-hand side vanishes near the boundary, the left hand side must vanish also.
This can happen in two possible ways: as $\l\to 0$
either $\xi\to 0$ or $\xi\to\xi_*\not=0$, with $F_{\xi}(\xi_*)=0$.
We will examine both options in turn and show that only the second one can consistently  happen.

\subsection{Vanishing $\xi$ in the UV}

If $\xi\to 0$, as $\l\to 0$  then $\xi$ must vanish as\footnote{This assumes that $\xi F_{\xi}$ remains constant as $\xi\to 0$.
Otherwise it can be shown that there is no solution.}
\be
\xi\simeq {\l^2\over F^2_{\xi}(R_*,0)~\la^2}+{\cal O}(\l^4)
\ee

Assuming an AdS$_5$ (constant curvature) solution $R=R_{*}$ equations (\ref{b9})-(\ref{b11}) imply
to leading order ($\l=0$) that
\be
F(R_*,0)=F_{R}(R_*,0)=0
\label{b13}\ee
This in turn implies that the AdS curvature $R_*$ must be a double root of $F(R,0)$.

We now move to the next order and perturb the leading solution $R=R_{*}, \l=0$ to $R=R_*+\delta R$ and non-zero but small $\l$,
\be
\xi={\l^2\over F_{\xi}^2(R_*,0)~\la^2}\left[1-{F_{\xi\xi}(R_*,0)\over
F_{\xi}^3(R_*,0)}{\l^2\over\la^2}-2{F_{\xi R}(R_*,0)\over F_{\xi}(R_*,0)}\delta R+\cdots\right]
\label{b14}\ee
\be
F\simeq {\l^2\over  F_{\xi}(R_*,0)~\la^2}+\cdots \sp \xi F_{\xi}\simeq {\l^2\over  F_{\xi}(R_*,0)~\la^2}+\cdots
\label{b15}\ee
where we have used (\ref{b13}).
To investigate this next order solution we use the desired asymptotic form of the dilaton
\be
\l\simeq -{1\over b_0\log(\Lambda r)}+\cdots
~~~\to~~~\phi\simeq {\rm constant} -\log[-b_0\log(\Lambda r)]+\cdots
\label{b16}\ee
and using the AdS metric
\be
ds^2={\la^2\over r^2}[dr^2+dx^{\mu}dx_{\mu}]\sp R_*=-{20\over \la^2}
\label{b17}\ee
to leading order we obtain
\be
\square\phi\simeq {4\over \la^2\lo}+\cdots \sp
(\pa \phi)^2\simeq {1\over \la^2\lo^2}+\cdots
\label{b17a}\ee

Equations (\ref{b15}) and (\ref{b17}) are now incompatible with (\ref{b11}).
Therefore, the starting assumption cannot be correct for the string theory we are seeking\footnote{A more exotic possibility is to cancel the offending term
against an $\delta R^2$ term from the next order expansion of $F$. This however will give an effective expansion in powers of $\l^{1\over 2}$ and not integral
powers of $\l$, a fact  at odds with QCD perturbation theory.}.

\subsubsection{Non-vanishing $\xi$ in the UV\label{uuv}}

In this case the function $F_{\xi}$ must have a zero as a function of $\xi$ at a non-trivial value $\xi_*(R)$.
We can therefore parametrize for convenience
\be
F\simeq c_0(R)+{c_1(R)\over 2}(\xi-\xi_*(R))^2+{\cal O}\left[ (\xi-\xi_*(R))^3\right]
\ee
and obtain\footnote{For similar reasons
the more general possibility $F\simeq c_0(R)+{c_1(R)\over 2}(\xi-\xi_*(R))^{a+1}+\cdots $ with $a> 0$  would
imply $\l\sim \lo^{-{1\over a}}$. Therefore only the case $a=1$ is relevant for YM.}
\be
\xi\equiv \xi_*(R) +\delta\xi\simeq \xi_*(R) \pm {\l\over c_1(R)~\la~\sqrt{\xi_*(R)}}+{\cal O}(\l^2)
\label{b18}\ee
We keep both signs here but below we will see that only the minus sign is relevant.

Again the gravitational equation implies that for  AdS to be the leading solution (at $\l=0$) we must have
\be
c_0(R_*)=0\sp {\pa c_0\over \partial R}\Big |_{R=R_*}=0
\label{b19}\ee
Using the above we  obtain that $F$ is zero to next order and the first non-trivial contribution is at quadratic order
\be
F(R,\xi)={\l^2\over 2c_1(R_*)~\la^2~\xi_*(R_*)}+{1\over 2}{\pa^2 c_0\over \pa R^2}(R_*)(R-R_*)^2+\cdots
\label{b20}\ee
and therefore subleading while
\be
\xi F_{\xi}=\pm\sqrt{\xi_*(R_*)}~{\l\over \la}+\cdots
\label{b21}\ee
Now it is possible to solve (\ref{b11}) to leading order and match
\be
b_0={\la \sqrt{\xi_*(R_*)}\over 16}
\label{b22}\ee
To have asymptotic freedom  we must choose the minus sign in (\ref{b18}).

Continuing further to the trace of the gravitational equation (\ref{b10}) we obtain
\be
F_R~R+4\square F_R-16\pa \phi\pa F_R-8F_R\square\phi+10(\pa\phi)^2-{5\over 2}F+5\xi F_{\xi}=0
\label{b23}\ee
Taking into account that to leading order $F_R=0$, to next order $F=0$ and (\ref{b19}) we obtain
that (\ref{b23}) becomes to next order,
\be
(4\square +R_*)\delta R={5+{{\delta \xi_*\over \delta R}(R*)\over \xi_*(R_*)}R_*\over c_0''(R_*)}2\sqrt{\xi_*(R_*)}{\l\over \la}
\label{b25}\ee
where we have used the fact $\square \l=-{4\over b_0\la^2\lo^2}$ which gives a subleading contribution.
This equation gives the following leading modification  to the AdS$_5$ metric
\be
b={\ell\over r}\left[1+{w\over \lo}+\cdots\right]
\sp \delta R={40w\over \ell^2\lo}+\cdots
\label{b26}\ee
with
\be
w={-5+{{\delta \xi_*\over \delta R}(R*)\over \xi_*(R_*)}R_*\over c_0''(R_*) }{\xi_*(R_*)\over 80R_*}
\label{b27}\ee

Let us pause and review what we have found. Near the boundary the leading solution is AdS$_5$ and the dilaton is such that
$\l=0$.

\begin{itemize}

\item  As all dilaton-dependent parts of the effective action  (\ref{b9}) are subleading near the boundary, the leading AdS$_5$
solution must be supported by curvature alone. This is the essence of the conditions (\ref{b19}).
In particular the non-critical dilaton potential and the corrections coming from integrating out the four-form
 are subleading near the boundary due to asymptotic freedom.

The fact that the AdS solution is supported by curvature alone implies
that modulo accidents, the AdS curvature scale $\la$ is of the same order of magnitude as the
the string scale $\ls$.
Do we know string backgrounds with such a property? In a sense yes,  although the  backgrounds we know are somewhat simpler.
In particular  coset CFTs, \cite{tseytlin2} share some features with what should happen here. In coset CFTs there is a special frame
in which the solution to the $\sigma$ model conditions can be thought as exact, but in other schemes it obtains corrections.
However, unlike what is expected to happen here, in coset models one can vary (in a discrete fashion)
 the curvatures by varying the levels of the current algebra. In a sense we must have a solution where the curvature cannot be varied.

\item  Unlike situations in critical string theory, the asymptotic AdS geometry here is not supported by RR flux.
 Therefore, to leading order near the boundary the $\sigma$-model is ``conventional. It is to next order that the flux enters the
 solution and the coupling starts to run.

\item  So far we have seen the importance of non-linearity of the curvature part of the effective string action in order to find
an leading AdS solution.
 Note however that in order for the function $F_{\xi}$ to have a root at a non-zero value of $\xi$, the function $F(\xi)$ must also be non-linear.
Therefore, we also need the higher derivative corrections of the four-form to find the asymptotic solution of YM.

\item  Many authors have suggested that the key to understanding the UV region of the QCD string theory is a theory of an infinite number of massless
higher spin fields. The idea behind is coming from ${\cal N}=4$ holographic intuition.
There, $\l\to 0$ implies that $\a'\to\infty$ and the whole string spectrum is becoming massless.
We see here that things work differently. The string has finite stiffness at the boundary, although it is soft, with a
fundamental string tension that is of the same order of magnitude as the background curvature.
On the other hand, all three and higher-point connected correlators will vanish near the boundary for a simple reason.
When properly normalized,  they
are all multiplied with positive powers of $\l$ that vanishes near the boundary.

\item In this section we worked out explicitly the first non-trivial order of the string equations.
We have ``imposed" both an asymptotic AdS solution and a leading running of the coupling constant.
For the  rest,  assuming genericity,  there is a regular perturbation theory in inverse powers of logs, that
is similar to the perturbative expansion in perturbative QCD. The various coefficients arise from the
expansion coefficients of the non-linear string effective action around the vacuum solution.
Without complete control of the non-critical string theory they cannot be calculated.
Despite this, the structure of the near-boundary perturbation theory is clear. Moreover the simplifying assumptions we made
about the effective action do not seem to modify the conclusions above.

\end{itemize}

\section{The IR regime}

We have seen that the general structure of the string effective action and some simple assumptions on the UV asymptotics
indicate the presence of the standard YM perturbation theory in the UV.
The situation in the IR is much murkier for a the simple reason, that no guiding principle like perturbation theory is known to exist.
We do have specific expectations however from the strong coupling region of QCD.
In particular we expect confinement and a discrete and gapped glueball spectrum.
However these requirements are fairly indirect to guide us in the IR.

A first question we would like to ponder is what happens to the dilaton in the IR.
The most natural expectation is that increases without bound, so that $\l\to\infty$.
There have been minority claims of an IR fixed point in QCD but such claims are not in our opinion credible.
There is also the alternative that the coupling asymptotes to a finite large value in the IR.
Although this is not excluded, we will not entertain it here for two reasons. The first is that we did not find a good way to
implement this possibility while keeping all the properties we expect from YM in the IR, in particular confinement.
 The second is that even if $\l\to\infty$ in the IR for most observables ,
there may be  a maximum finite value for the 't Hooft coupling if all low lying wave-functions have support away from the $\l\to\infty$ limit.

Vacua with a runaway dilaton are known in string theory. The simplest
 is the linear dilaton vacuum that is simple enough to define,
but has consistently puzzled researchers for two decades. Due to advances
 in the understanding of Liouville theory, the associated matrix models
and more recent advances in the associated world-sheet CFTs we have today
 a fairly good idea of the physics in such a background.
A crucial ingredient is that there should be a sufficient  screening
of the strong coupling singularity,
 that in Liouville theory is achieved via the ``Liouville wall".
 We will see later on that a linear dilaton background in the IR is
 marginally compatible with what we expect from YM  at strong coupling.
 In particular we can show that we have confinement and a mass gap,
  but the spectrum is continuous above the gap. However, by slightly modifying the
  background we will obtain also a discrete spectrum as well as linear
   asymptotic trajectories.
  Moreover as will see,  there is a sense in which the strong coupling
  singularity is screened: all local low-energy observables do not get
  contributions from arbitrarily close to the singularity. This is what
   we will call a ``repulsive" singularity that
  is a more constrained concept than the ``good" singularities of Gubser, \cite{gubserbad}.

Another intuition is emerging from the ${\cal N}=4$ paradigm of AdS/CFT:
 at strong coupling we could expect a good effective description
 in terms of a two-derivative action. Although this is rather transparent
  in standard AdS/CFT, it is less clear here,
 because the 't Hooft coupling is not constant. We will however take
 it as a principle and we will see how far we can go.
 An important ingredient in order to implement this idea is that the
 curvature in the string frame must be very small in the IR.
 This ties well together with the linear dilaton paradigm as in that
  case the curvature in the string frame vanishes.
 We would therefore investigate the possibility that there is a vacuum
  solution in the IR with $\l\to \infty$, a small curvature in the string frame,
 and subleading contributions  from higher derivative terms.

  To investigate this we will expand now the string effective action in (\ref{b9a}) in powers of the curvature, remembering
  that $\xi$ is an implicit function of $R$ and $\l$ from (\ref{b5})
\be
F(R,\xi)-2\xi F_{\xi}(R,\xi)\equiv \sum_{n=0}^{\infty}Z_{n}(\l)~R^n
\label{b30}\ee
so that the action becomes
\be
S_{\rm tree}=\int d^5x\sqrt{g}e^{-2\phi}\left[4(\partial \phi)^2+ Z_0(\l)+Z_1(\l)R+ \sum_{n=2}^{\infty}Z_{n}(\l)~R^n\right]
\label{b31}
\ee
We now make a conformal transformation
\be
g_{\m\n}\to f(\phi)g_{\m\n}\sp R\to {1\over f}\left[R-4\square \log f-3(\partial \log f)^2\right]
\sp \square \to {1\over f}\left[\square +{3\over 2}\nabla^{\m}(\log f)\nabla_{\m}\right]
\label{b32}\ee
with
\be
f=e^{{4\over 3}\phi}Z_1^{-2/3}
\label{b34}\ee
to obtain
\be
S_{\rm tree}=\int d^5x\sqrt{g}\left[R-{4\over 3}\left[4-3{(1+Z_1')\over Z_1}\right]
{(\partial \phi)^2}+ {e^{4\phi\over 3}Z_0\over Z_1^{5\over 3}}+ \sum_{n=2}^{\infty}{Z_{n}(\l)Z_1^{{1\over 3}(2n-5)}\over e^{{4\over 3}\phi(n-1)}}~
\left(R+.....\right)^n\right]
\label{b35}
\ee
We may now define a new scalar $\Phi$ with a canonical kinetic term
\be
{d\Phi\over d\phi}=\sqrt{4-3{(1+Z_1')\over Z_1}}
\label{b36}\ee
Asking for positivity inside the square root, severely constraints on how $Z_1$ depends on $\phi$.
In particular we find that
\be
Z_1(\phi)=C_1 e^{{4\over 3}\phi}+{\rm subleading}\sp {\rm as}~~~~\phi\to\infty.
\label{b37}\ee

The action (\ref{b35}) can now be written as
\be
S_{\rm tree}=\int d^5x\sqrt{g}\left[R-{4\over 3}
{(\partial \Phi)^2}+ V_0(\Phi)+ \sum_{n=2}^{\infty}V_n(\Phi)~
\left(R+.....\right)^n\right]
\label{b35a}
\ee
with
\be
 V_0(\Phi)={e^{4\phi\over 3}Z_0\over Z_1^{5\over 3}}\sp V_n(\Phi)={Z_{n}(\l)Z_1^{{1\over 3}(2n-5)}\over e^{{4\over 3}\phi(n-1)}}
 \ee
 We may now return to a ``string frame" for $\Phi$ as $g_{\m\n}\to e^{-{4\over 3}\Phi}g_{\m\n}$ to obtain

 \be
S_{\rm ms}=\int d^5x\sqrt{g}e^{-2\Phi}\left[R+4
{(\partial \Phi)^2}+ W_0(\Phi)+ \sum_{n=2}^{\infty}W_n(\Phi)~{\cal R}^n\right]
\label{b35as}
\ee
where
\be
W_0(\Phi)={Z_0\over Z_1}~{1\over F^{2\over 3}}\sp W_n(\Phi)={Z_{n}\over Z_1}~F^{2(n-1)\over 3}\sp F=Z_1~e^{2(\Phi-\phi)}
\ee
\be
{\cal R}
=R+{8\over 3}\square\log F-{4\over 3}(\pa\log F)^2
\ee
From now on we will always call $\l=e^{\Phi}$ the 't Hooft coupling and use it interchangeably with $\Phi$.

Before we proceed further we should discuss our expectations on the dependence of the coefficients $Z_n$ on the 't Hooft coupling in the IR.
We do expect the leading dependence to be the same for all $Z_n$. Moreover, we expect that their growth in the IR is bounded.
This is already suggested by the positivity bound on $Z_1$ that we pointed out in (\ref{b37}).

In \cite{ihqcd2} an analysis of various dilaton potentials was performed, assuming a two-derivative action,
in order to find which ones give properties that we expect from YM, namely, confinement, mass gap and discrete spectrum.
At a two-derivative level, positive energy implies that all Lorentz invariant
ansatze lead to an IR singularity  (the only other alternative is an IR AdS
space). Therefore part of the criteria we used included that this singularity is
repulsive\footnote{This is a stronger condition than the one in Gubser's classification, \cite{gubserbad}.}:
spectra can be computed without the need of extra boundary conditions at the singularity and
string world-sheets should not come very close to the singularity.

It was found that potentials fall in several categories, that we parametrize their asymptotics as $V\sim \l^Q (\log \l)^P$.
The presence of the log terms matter only when $Q={4\over 3}$.
\begin{itemize}

\item Potentials with $Q<{4\over 3}$ do not confine.

\item Potentials of the form  $V\sim \l^{4/3}(\log\l)^{a-1\over a}$, $a>0$, satisfy all criteria,
and moreover they have the property that the string frame curvature vanishes in the IR.
For $a=2$ they generate asymptotically linear trajectories.
$a=1$ is the marginal case of the linear dilaton vacuum. This case does not have a discrete spectrum.

\item Potentials with $Q=4/3$ and $P>1$, confine properly, but the string frame curvature blows up at the singularity

\item Potentials with ${4\over 3}\leq Q\leq {4\sqrt{2}\over 3}$ confine properly  but the string frame curvature blows up at the singularity.

\item Potentials with $Q\geq {4\sqrt{2}\over 3}$ have a ``bad" (non-repulsive) IR singularity.

\end{itemize}

In \cite{ihqcd2} a detailed analysis of the glueball spectra of different confining potentials were performed.
It was found that only the ``soft potentials"  $V\sim \l^{4/3}(\log\l)^{a-1\over a}$, $a>0$ are
 capable of giving spectra that are reasonably close to the lattice glueball spectra
All other potentials produced glueball splittings (after adjusting/fitting parameters) that cannot be accommodated by the lattice data.

In view of this we may analyze the possibilities in the IR using the actions (\ref{b35a}) and (\ref{b35as}) and the principles that $Z_n/Z_1\to $ constant
as $\l\to\infty$ and that the bound (\ref{b37}) applies.

We find the following possibilities

\begin{enumerate}

\item When $Z_1\sim e^{{4\over 3}\phi}$ we find a potential that leads to non-confining behavior, and higher derivative corrections that are no controlable.

\item When $Z_1\sim e^{b\phi}\phi^c$ with $b<{4\over 3}$ leads to non-controlable higher derivatives with the exception of $b=-4+2\sqrt{7}\simeq 1.29$.
In the latter case the potential is $V_0\sim e^{{4\over 3}\Phi}\Phi^{{4\over 3}c}$ and this is an good confining potential with vanishing string frame
curvature provided $c<{3\over 4}$. In this case all higher derivative corrections are subleading.

\item When $Z_1\sim $ constant plus subleading we end up with a non-controlable higher derivative behavior with the exception of
$Z_1=1-{C\over 2\phi}+\cdots$. In that case the leading effective potential is $V\sim e^{{4\over 3}\Phi}\Phi^{C}$ which gives rise to confining behavior
and vanishing string frame curvature. In this case the higher derivative corrections are again subleading.

\item Any other asymptotic behavior leads to non-controlable higher-derivative corrections.

\end{enumerate}

It is not obvious which of the two favorable possibilities found above are realized in QCD.
We cannot even exclude the cases that have non-controlable higher derivative corrections.
It is noteworthy though that both favorable cases 2 and 3, give the same order of magnitude corrections to the IR running of $\lambda$.
Starting from the  Einstein frame effective action and keeping the leading quadratic term (that is $\square \log F$)
we can directly estimate that its influence on the leading order solution scales as
\be
{\delta\l\over \l}\sim {1\over (\log \l)^2}
\ee
for all soft potentials $V\sim \l^{4\over 3}\l^{a-a\over a}$, $a>1$.

With a bit of optimism and guided by the intuition above
we will accept that the IR of QCD is governed by a soft potential, that once taking the existence of linear trajectories into account, it should
have the following asymptotics $V\sim \l^{4\over 3}\sqrt{\log \l}$. Moreover, in such a case the higher-derivative corrections are suppressed.

\subsection{The 't Hooft coupling in the IR}

We have already seen that in the IR, the canonically normalized dilaton field $\Phi$ is non-linearly related to the 
original string theory dilaton $\phi$. 
In particular in case 2 of this section with $b=-4+2\sqrt{7}$ this relation is 
\be
\Phi=(3-\sqrt{7})\phi+c\log \phi+\cdots
\ee
where $c$ a constant, while for case 3
\be
\Phi=\phi+c\log\phi+\cdots
\ee
where $c$ is again a constant. 

There is however another source of corrections to the identification 
of the 't Hooft coupling, \cite{ihqcd1} if it defined using the $D_3$ brane world-volume action
\footnote{We thank K. Kajantie whose question 
prompted this discussion.}.
The general form of the kinetic term for the gauge fields on a $D_3$ brane is expected to be
\be
{\cal L}_{F^2}=e^{-\phi}Z(R,\xi)~Tr[F^2]
\ee
where $Z(R,\xi)$ is an (unknown) function of curvatures and the five-form field strength.
At weak background fields $Z\simeq -{1\over 4}+\cdots$. 

In the UV regime, using the formulae of section \ref{uuv} we obtain

\be
{\cal L}_{F^2}=N_c~Tr[F^2]~{1\over \lambda}\left[Z(R_*,\xi_*)-{Z_{\xi}(R_*,\xi_*)\over F_{\xi\xi}(R_*,\xi_*)\sqrt{\xi_*}}{\l\over \la}+{\cal O}(\l^2)\right]
\ee
from which we can identify the QCD 't Hooft coupling as 
\be
\l_{QCD}={\l\over Z(R_*,\xi_*)}+{Z_{\xi}(R_*,\xi_*)\over Z(R_*,\xi_*)^2 F_{\xi\xi}(R_*,\xi_*)\sqrt{\xi_*}}{\l^2\over \la}+{\cal O}(\l^3)
\ee
indicating that there are ``perturbative" corrections to the identification to the 't Hooft coupling.
In the IR, the relation is more complicated and to leading order in the string curvature it becomes
\be
\l_{QCD}={\l\over Z(0,\xi_*(0,\l)}
\ee
 We do not know what this non-trivial function of the dilaton is.
 We should stress though that this is a particular definition (scheme) for the 't Hooft coupling constant (that we call the $D_3$ scheme,
 and it may be very different (especially in the IR) from other schemes used in lattice calculations, \cite{lat}.
   
   \section{Schemes and scheme-dependence}
As mentioned in the previous section, there are several choices in the theory that amount to a choice of scheme.
Schemes in QFT correspond to coordinate choices in the space of couplings. One can parametrize the space of couplings
in different fashions but that is not affecting physical observables that should ``coordinate invariant". A particular scheme choice 
is that of the renormalization group scale used in renormalizing the theory. Here again physical observables should be 
RG-invariant.

In the holographic context scheme dependence related to coupling redefinitions translates into field redefinitions for the bulk fields.
As the bulk theory is on-shell, all on-shell observables (that are evaluated at the single boundary of spacetime) are independent
of the field redefinitions showing the scheme-independence is expected.
Invariance under radial reparametrizations of scalar bulk invariants is equivalent to RG invariance.
Because of renormalization effects, the boundary is typically shifted and in this case field redefinitions must be combined
with appropriate radial diffeomorphisms that amount to RG-transformations.

In particular, the holographic definition of the ``Energy" should not affect physical quantities, provided it is reasonable: (a) it is monotonic 
and (b) it vanishes at the ultimate IR. Of course quantities like the $\beta$-function do depend on the
 definition of the energy (as well as the definition of the coupling).  This is expected as a $\beta$-function is not an invariant (physical)
 but a vector in coupling constant space, as well as in energy space and therefore changes in different schemes.

\section{A phenomenological model\label{pheno}}

In view of the discussion in the previous sections, we should ask the question: to what extend a tractable
simplified model can be devised that is reasonably close to what we expect to happen in the low-energy theory.

The answer to this question in the IR is straightforward. We have argued that the leading IR asymptotics are described by a dilaton gravity system
with a soft potential, and no higher-derivative corrections.
\footnote{Dilaton gravity with a potential was also argued to have similar thermodynamic properties to QCD in \cite{gubser}.}
\be
S_{HQCD}=M_p^3N_c^2\int d^5x\sqrt{g}\left[R-{4\over 3}
{(\partial \l)^2\over \l^2}+ V_0(\l)\right]\sp V_0\sim \l^{4\over 3}\sqrt{\log\l}~~~{\rm as}~~~\l\to \infty
\label{b55a}\ee

The non-trivial issue is the UV. There we have seen that higher-derivative corrections, especially in the curvature,
are essential for the appearance of the asymptotically AdS$_5$ solution and
a logarithmically-vanishing coupling constant. Implementing this at the present stage seems out of reach if we insist in a model
 that we can calculate observables.
There is short-cut though that has the advantage that it can capture some of the salient features of the UV solution and dynamics
while it remains manageable. As expected,  it has also shortcomings that we will list below.

The shortcut is to assume a similar action as in (\ref{b55a}) but with a potential that has different weak coupling asymptotics.
\be
V_0(\l)={12\over \la^2}\left[1+v_1\l+v_2\l^2+v_3\l^3+\cdots\right]~~~{\rm as}~~~\l\to 0
\label{b56}\ee
This form is indicated by the fact that if $v_1\not=0$ (that implies that
the ultimate boundary is not a minimum of the potential)
and the dilaton boundary conditions are carefully chosen (as detailed in
\cite{ihqcd2,ihqcd4}), the vacuum solution for the 't Hooft coupling and the metric
has the same structure in terms of the expansion in inverse logs, as the
one advocated here in section  \ref{uv} and which agrees with QCD perturbation theory.
In particular once we chose a definition of the energy
scale\footnote{In \cite{ihqcd1,ihqcd2} it was argued that a good (but not unique)
choice is the scale factor of the metric
in the Einstein frame. Moreover different choices that are perturbatively
related affect the $\beta$-function coefficients beyond the two-loop level.}, then the dimensionless coefficients
$v_n$ are in one to one correspondence with the $\beta$-function coefficients.

Moreover the basic property that as we reach the boundary three- and higher-point connected correlators vanish
is satisfied, since $\l\to 0$.

There are further consistency checks that such an approach in the UV captures quite a few important properties of the system.
We list them below
\begin{itemize}

\item It captures the correct number of UV boundary conditions in accordance with YM

\item It gives the correct perturbative running of the coupling once the analogue of $\Lambda_{QCD}$ is set. This was verified in
\cite{ihqcd2} by reproduced the experimentally measured value of the strong coupling constant at $E=1.2$ GeV, using as an input the lowest glueball mass.

\item It provides the correct UV boundary conditions for the physical fluctuations of the system, the glueballs, \cite{nitti,ihqcd2}.

\item It is compatible with the existence of a finite temperature phase with the correct physical properties,  \cite{ihqcd3,ihqcd4}.

\item It provides the correct free-field limit at high temperature and its first logarithmic correction.

\end{itemize}

\noindent
There are also issues that are not as they should be. They include

\begin{itemize}

\item The conformal anomaly. It is well known \cite{skenderis} that any theory with a leading two-derivative bulk description
must be such that the two anomaly coefficients are equal $a=c$ to leading order, ${\cal O}(N_c^2)$.
On the other hand, QCD with or without quarks, has $a\not= c$ to leading order in $N_c$\footnote{It can be shown the for CFTs, with a minimum
of ${\cal N}=1$ supersymmetry only fundamental matter can contribute to $a-c$, \cite{bcckp}. For theories with a weak coupling limit, conformal invariance implies that
only theories with the ${\cal N}=4$ spectrum have $a=c$.}.

\item The shear viscosity. It was shown in \cite{buliu} that two-derivative
theories of gravity coupled to matter, have all the minimum constant value of
the $\eta\over s$ ratio.
For the IR regime, as we argued earlier this is probably a very good approximation.
However as we approach the UV, this fact is more and more at odds with what happens in perturbative QCD
where $\eta\over s$ diverges as the coupling vanishes.

\item In the UV there are some spurious extra logs in some quantities. They reflect the fact that in the string theory dual of YM,
the asymptotic AdS metric is achieved in the string  frame and not the Einstein frame. This is a non-trivial issue and can be
ascertained from various points of view. The analysis in section \ref{eft} is performed in the string frame. It would not go through as
such if an AdS metric in the Einstein frame is advocated. On the other hand, for the phenomenological model discussed in this section, it is not consistent
with the equations of motion to impose the AdS metric in the string frame. We therefore impose it on the Einstein frame. The difference is an extra multiplicative
factor of $(\log(\Lambda r))^{4\over 3}$ since the dilaton is asymptotically logarithmic.
The effects of this result in mild differences in some quantities. For example a calculation
of the short distance asymptotics of the static quark potential produces
an an  extra log factor  of the distance,  multiplying the $1/r$ of the Coulomb law. This modification is mild and was
shown to fit as well baryonium data as the standard Cornell potential,  \cite{zeng}.

\item Higher order correlators of the basic operators of the theory,
 or higher perturbative (ie. log corrections) to low-order correlators may be quantitatively different although generically they remain qualitatively correct.

\end{itemize}

\subsection{Parameters, fits and predictions}

The action (\ref{b55a}) contains as parameters two scales,  $M_p$ and $\la$ from the potential (\ref{b56})
 as well as in principle many dimensionless parameters hidden in the potential.
 There is also the string scale $\ls$ that enters the fundamental string action.
 We do not know ab initio how such scales are related.
 However we may observe that $\la$ is not really a physical parameter by a choice of energy scale.

 Moreover there is dependence on boundary conditions for the vacuum solution. There are three boundary conditions needed
 for the gravity-dilaton system. One of them is a gauge artifact, as it can be identified with a translation of origin in the radial coordinate.
 Another is fixed by demanding that correct perturbative asymptotics of the dilaton and the absence of a bad singularity in the bulk.
 The third corresponds to $\Lambda_{QCD}$. It is defined in a reparametrization invariant way as
 \be \label{Lambda}
\Lambda = \ell^{-1} \lim_{\l\to 0} \left\{b(\l){\exp\left[ - {1\over b_0\l}\right] \over \l^{b_1/b_0^2}} \right\},
\ee
where $b(\l)$ is the scale factor of the Einstein metric in (\ref{g}) written as a function
 of $\l$, and $b_0,b_1$ are the one and two-loop coefficients of the
perturbative YM  $\beta$-function.
$\Lambda$ is fixed once we specify the value of the scale factor $b(\l)$ at a given $\l_0$.

 $M_p$ does not affect spectra of particles but it affects interactions. It also affects the size of the free-energy at finite temperature.
 It can be fixed,
 \be
 (M_p\la)^{-3} =45~\pi^2\;,
 \ee by demanding that the free energy for $T\to\infty$ asymptotes to that of a free gas of gluons.
 The fundamental string scale can be fixed by calculating the inter-quark
 potential and comparing with the associated lattice results for the effective string tension.

 In practice the rest of hidden parameters of the potential are truncated to a small set. In\cite{ihqcd2} a one-parameter fit could reproduce all known lattice
 glueball spectra with values inside the error bars, and all the characteristic  features of the thermodynamics, \cite{ihqcd3}.
 Adding a second parameter the full equilibrium thermodynamics fits perfectly well, \cite{to}.

 Once the parameters have been fixed, then one can calculate other quantities.
 They include the full spectrum \cite{ihqcd2}, two and higher-point functions,
 the full equilibrium thermodynamic quantities \cite{to}, as well as transport data.

 \addcontentsline{toc}{section}{Acknowledgements}
\section{Acknowledgements}

The author would like to thank Umut Gursoy, Fransesco Nitti and Liuba Mazzanti
for an enjoyable collaboration on the general subject of this paper.
He would like to also thank numerous colleagues that have shared intuition knowledge and criticism that has been constructive.
In particular discussions with  Ofer Aharony, Luis Alvarez-Gaum\'e,   Richard Brower,  Frank Ferrari,  Steve Gubser,
Gary Horowitz, Thomas Hertog, Edmond Iancu, Frithjof Karsch, David  Kutasov, Hong Liu,
 Carlos Nunez,  Andrei Parnachev, Ioannis Papadimitriu, Krishna Rajagopal,
Edward Shuryak,
Kostas Skenderis, Dam Son, Jacob Sonnenschein,  Shigeki Sugimoto, Marika Taylor  and Lawrence Yaffe are gratefully acknowledged.

The kind invitations to lecture and hospitality at Moriond, the string theory workshop at the Institute for Advanced Studies
in Jerushalem, the IPM school in Isfahan, Iran, the Galileo Galilei Institute program ``From lattice to AdS/CFT",
the ``4th RTN-EU workshop" in Varna, Bulgaria, the CERN Institute on black holes, and the Superstrings\@ Cyprus meeting of the String theory network
are acknowledged.

This work was partially supported by ANR grant NT05-1-41861,
RTN contracts MRTN-CT-2004-005104 and MRTN-CT-2004-503369,
CNRS PICS  3059 and 3747 and by a European Excellence Grant,
MEXT-CT-2003-509661.

\end{document}